\documentclass[12pt]{article}
\usepackage{axodraw,bbold}

\catcode`@=12
\begin{document}
\thispagestyle{empty}
\begin{flushright} 
UCRHEP-T449\\ 
March 2008\
\end{flushright}
\vspace{0.5in}
\begin{center}
{\LARGE	\bf B and not L in Supersymmetry:\\New U(1) Gauge symmetry\\
and Dark Matter\\}
\vspace{1.0in}
{\bf Ernest Ma\\}
\vspace{0.2in}
{\sl Department of Physics and Astronomy, University of California,\\}
\vspace{0.1in}
{\sl Riverside, California 92521, USA\\}
\vspace{1.5in}
\end{center}

\begin{abstract}\
To \emph{enforce} the conservation of baryon number $B$ and not lepton 
number $L$ in supersymmetry, a new $U(1)_X$ gauge symmetry is recommended. 
An example is offered with new particles interacting under $U(1)_X$ 
which are good candidates for the dark matter of the Universe.
\end{abstract}

\newpage
\baselineskip 24pt

In an unrestricted supersymmetric extension of the Standard Model (SM) of 
particle interactions, there are the following well-known allowed bilinear 
and trilinear terms:
\begin{equation}
L_i \Phi_2, ~~ L_i L_j e^c_k, ~~ L_i Q_j d^c_k, ~~ u^c_i d^c_j d^c_k,
\end{equation}
where $L_i = (\nu_i,e_i)$, $Q_i = (u_i,d_i)$, $\Phi_1 = (\phi^0_1,\phi^-_1)$, 
$\Phi_2 = (\phi^+_2,\phi^0_2)$, etc.  This is of course unacceptable because 
both baryon number $B$ and lepton number $L$ are not conserved, and rapid 
proton decay cannot be avoided.  The conventional choice of the Minimal 
Supersymmetric Standard Model (MSSM) is to impose by hand the notion of 
$R$ parity for each particle, which is just the product of its multiplicative 
baryon number $(-)^{3B}$, its multiplicative lepton number $(-)^L$, and its 
intrinsic spin parity $(-)^{2j}$.  As a result, all four terms of Eq.~(1) 
are forbidden.  This has the desirable consequence of an absolutely stable 
particle odd under $R$ which is a candidate for the dark matter of the 
Universe.

Another choice is to forbid only the last term of Eq.~(1) by hand, thereby 
conserving $B$, and allow the other three terms, thereby violating $L$.  
This is acceptable because $B$ conservation by itself is sufficient to 
forbid proton decay.  Such models of $R$ parity violation have been 
discussed extensively in the literature.  Of course, there is no longer 
any dark-matter candidate, and no better understanding as to why $B$ is 
conserved and not $L$.  In any case, whether $R$ parity is conserved or not, 
there remains the puzzle of the allowed bilinear term $\mu \Phi_1 \Phi_2$.  
It is not understood why $\mu$ should be of order TeV or less, and not some 
much larger fundamental scale, as would be expected. 

In this note, a new $U(1)_X$ gauge symmetry is proposed, whereby $B$ is 
conserved but not $L$, and the scale of $\mu$ is determined by the 
spontaneous breaking of $U(1)_X$.  In addition, new particles exist which 
are good dark-matter candidates.  The idea of using a particular new U(1) 
gauge symmetry to explain the $\mu$ puzzle and to prevent proton decay is 
not new \cite{cdm98,e00,ao00,m02}.  Recently, it has also been applied to 
enforcing either $B$ conservation or $L$ conservation or both \cite{lmw08}.  
Here, the proposal is to conserve $B$ and not $L$, in keeping with most 
phenomenological studies of $R$ parity violation, \emph{and to accommodate 
dark matter}.

Consider first the particles of the MSSM and their transformations under 
$U(1)_X$ as shown in Table 1, where \emph{three} families of quarks and 
leptons are understood.  The terms $Q u^c \Phi_2$ and $Q d^c \Phi_1$ 
are already allowed.  To have $L e^c \Phi_1$ as well as $L L e^c$ and 
$L Q d^c$ terms, $n_4 = -n_1 -n_3$ and $n_5 = 2n_1 + 2n_3$ are needed. 
To forbid $u^c d^c d^c$, the condition $n_2 + 2n_3 \neq 0$ is required.

\begin{table}[htb]
\caption{MSSM particle content of proposed model.}
\begin{center}
\begin{tabular}{|c|c|c|}
\hline 
Superfield & $SU(3)_C \times SU(2)_L \times U(1)_Y$ & $U(1)_X$ \\ 
\hline
$Q \equiv (u,d)$ & $(3,2,1/6)$ & $n_1$ \\ 
$u^c$ & $(3^*,1,-2/3)$ & $n_2$ \\ 
$d^c$ & $(3^*,1,1/3)$ & $n_3$ \\ 
$L \equiv (\nu,e)$ & $(1,2,-1/2)$ & $n_4$ \\ 
$e^c$ & $(1,1,1)$ & $n_5$ \\ 
\hline
$\Phi_1 \equiv (\phi^0_1,\phi^-_1)$ & $(1,2,-1/2)$ & $-n_1-n_3$ \\ 
$\Phi_2 \equiv (\phi^+_2,\phi^0_2)$ & $(1,2,1/2)$ & $-n_1-n_2$ \\ 
\hline
\end{tabular}
\end{center}
\end{table}

Consider the addition of a pair of color-triplet superfields $(h,h^c)$ and 
one electroweak triplet superfield $\Sigma = (\Sigma^+,\Sigma^0,\Sigma^-)$.
\begin{eqnarray}
h &\sim& (3,1,-1/3,n_6),\\
h^c &\sim& (3^*,1,1/3,n_7),\\
\Sigma &\sim& (1,3,0,n_8).
\end{eqnarray}
The absence of the $[SU(3)]^2 U(1)_X$, $[SU(2)]^2 U(1)_X$, and $[U(1)_Y]^2 
U(1)_X$ anomalies requires
\begin{eqnarray}
&& 6n_1 + 3n_2 + 3n_3 + (n_6+n_7) = 0, \\
&& 4n_1 - n_2 - 4n_3 + 4n_8 = 0, \\
&& 24n_1 + 21n_2 + 30n_3 + 2(n_6+n_7) = 0.
\end{eqnarray}
As a result,
\begin{eqnarray}
&& n_1 = -{5 \over 4}n_2 - 2n_3, ~~~~ n_4 = {5 \over 4}n_2 + n_3, ~~~~ n_5 = 
-{5 \over 2}n_2 - 2n_3, \\
&& n_6+n_7 = {9 \over 2}(n_2+2n_3), ~~~~ n_8 = {3 \over 2}(n_2 + 2n_3).
\end{eqnarray}
The absence of the $U(1)_Y [U(1)_X]^2$ anomaly implies
\begin{eqnarray}
45 n_2^2 + 144 n_2 n_3 + 108 n_3^2 - 4  n_6^2 + 4 n_7^2 &=& \nonumber \\
9(n_2+2n_3)(5n_2 + 6n_3 - 2n_6 + 2n_7) &=& 0.
\end{eqnarray}
Hence $n_2+2n_3 \neq 0$ implies
\begin{equation}
n_6 = {7 \over 2}n_2 + 6n_3, ~~~~ n_7 = n_2 + 3n_3.
\end{equation}

The crucial condition $n_2+2n_3 \neq 0$ also forbids the 
trilinear terms $u^c d^c d^c$, $u^c d^c h^c$, $Q Q h$, and $L Q h^c$, as 
well as the bilinear terms $\Phi_1 \Phi_2$, $L \Phi_2$, and $d^c h$.  
On the other hand, the terms $Q u^c \Phi_2$, $Q d^c \Phi_1$, $L e^c \Phi_1$, 
$L L e^c$, and $L Q d^c$ are allowed, thus conserving $B$ but not $L$.

At this point, the model is incomplete because of the lack of mass terms 
for $\Phi_1 \Phi_2$, $h h^c$, and $\Sigma \Sigma$.  Singlet superfields 
$\chi_{3,6,9}$ are then required, transforming under $U(1)_X$ as $-3,-6,-9$ 
respectively in units of $(n_2+2n_3)/2$.  To allow the exotic $h,h^c$ quarks 
to decay, another $\chi_7 \sim -7$ is needed so that $d^c h \chi_7$ is 
possible. With this particle content, the sum of $U(1)_X$ charges is 
$-5(n_2+2n_3)$ and the sum of $[U(1)_X]^3$ charges is $-80(n_2+2n_3)^3$.  
To cancel these anomalies, the following singlets may be added: one copy 
each of $\chi_1 \sim -1$, $\chi_4 \sim -4$, $\chi_{10} \sim 10$, three copies 
of $\chi_5 \sim -5$, and ten copies of $\chi_2 \sim 2$, again in units of 
$(n_2+n_3)/2$.  Note that these singlets are chosen so that there are no 
bilinear terms (i.e. no two with opposite charges), otherewise the analog 
of a $\mu$ term would be allowed, thereby defeating the purpose of having 
a new $U(1)_X$ gauge symmetry to forbid such terms in the first place.  
Allowed trilinear terms are $\chi_2 \chi_2 \chi_4$, $\chi_1 \chi_9 \chi_{10}$, 
$\chi_3 \chi_7 \chi_{10}$, $\chi_4 \chi_6 \chi_{10}$, and $\chi_5 \chi_5 
\chi_{10}$.  Since their scalar components may all have nonzero vacuum 
expectation values, their fermion components all become massive at that 
energy scale.  Thus an explicit and completely consistent example 
exists for an anomly-free $U(1)_X$ which conserves $B$ but not $L$.

As a byproduct of $U(1)_X$, dark-matter candidates also exist. For example, 
$\chi_2$ or $\chi_5$ may be assigned odd under an exactly conserved $Z_2$ 
symmetry, in which case they must also have zero vacuum expectation values.
They may annihilate in the early Universe through the $U(1)_X$ gauge boson 
into the usual quarks and leptons with a cross section characterized by 
the scale of $U(1)_X$ breaking, i.e. of order TeV.  Their elastc interaction 
with nuclei in direct-search experiments may be suppressed at the same 
time by choosing $n_3 = -3n_2/4$ so that the $U(1)_X$ coupling to an 
isoscalar combination of quarks is purely axial-vector.

An even better candidate for dark matter \cite{m05,cfs06} is the $\Sigma^0$ 
fermion, without which $n_2 + 2n_3 \neq 0$ cannot be realized.  The mass 
difference 
$m_{\Sigma^\pm} - m_{\Sigma^0}$ is given by
\begin{equation}
\Delta m_\Sigma = {\alpha \over \pi} {m_W^2 m_\Sigma \over \sin^2 \theta_W} 
\left[ {1 \over m_\Sigma^2-m_W^2} \ln {m_\Sigma^2 \over m_W^2} - 
{1 \over m_\Sigma^2-m_Z^2} \ln {m_\Sigma^2 \over m_Z^2} \right].
\end{equation}
This splitting is always positive and has a maximum of about 115 MeV 
for $m_\Sigma = 40$ GeV.  Hence the decay of $\Sigma^+$ will likely be 
exclusively into $\Sigma^0 e^+ \nu$.

In conclusion, it has been shown that $R$ parity violation, in the sense of 
$B$ conservation but not $L$, may be enforced by a new $U(1)_X$ gauge symmetry 
which also forbids all bilinear mass terms.  The scale of $U(1)_X$ breaking 
as well as electroweak symmetry breaking are then related to that of 
supersymmetry breaking.  A consistent example is presented, with new 
particles transforming under $U(1)_X$ as dark-matter candidates.

This work was supported in part by the U.~S.~Department of Energy under Grant 
No.~DE-FG03-94ER40837.

\bibliographystyle{unsrt}

\end{document}